\documentclass[preprint,bibnotes,showpacs,preprintnumbers,amsmath,amssymb,prd,superscriptaddress]{revtex4}  

\usepackage{color}

\usepackage{graphics}
\usepackage{dcolumn}% Align table columns on decimal point
\usepackage{bm}% bold mathtj
\usepackage{bm,amsmath}
\usepackage{multirow}
\usepackage{setspace}
\usepackage[dvips]{graphicx}
\usepackage{ascmac}
\usepackage{longtable}
\usepackage{ulem}  % To use \sout{} or \xout{}

\begin{document}
%
%\thesaurus{02(02.14.1; 12.03.4; 12.05.1)}

%%%%%%%%%%%%%%%%%%%%%%%%%%
\title{%Upper limit of quiescent X-ray luminosity from accreting neutron stars with  neutrino losses due to strong pion condensations
Thermal structures of accreting neutron stars with neutrino losses due to strong pion condensations}

\author{Y. Matsuo}
\affiliation{Department of Physics, Kyushu University,
Fukuoka 819-0395, Japan}

\author{M. Hashimoto\footnote{E-mail: hashimoto@phys.kyushu-u.ac.jp}}
\affiliation{Department of Physics, Kyushu University, 
Fukuoka 819-0395, Japan}

\author{K. Hayashida}
\affiliation{Department of Physics, Kyushu University, 
Fukuoka 819-0395, Japan}

\author{H. Liu}
\affiliation{Department of Physics, Kyushu University, 
Fukuoka 819-0395, Japan}

\author{T. Noda}
\affiliation{Kurume Institute of Technology, 
Fukuoka 830-0052,Japan}

\author{M. Y. Fujimoto}
\affiliation{Department of Physics, Hokkaido University, 
Sapporo 060-8810, Japan}

%\date{Received / Accepted  }

%%%
\begin{abstract}
%We construct the quiescent neutron star models in the evolutionary calculations without the compressional heating.
%The X-ray luminosities have been derived in terms of the time-averaged mass accretion rate for various neutron star masses and surface compositions. 
%We compare the quiescent luminosities observed from X-ray transients in low mass X-ray binaries,
%    where the stellar evolutionary calculations of accreting neutron stars include neutrino cooling due to strong pion condensations. 
%Since we neglect the compressional heating, obtained relations between the core and surface temperatures are consistent with those of previous studies.
%However, it is found that we do not need strong cooling processes such as direct Urca processes concerning nucleons and/or hyperons to explain the observations of SAX J1808.4-3658
%   if we use the neutrino emission rate due to strong pion condensation.

Quiescent X-ray luminosities are presented in low mass X-ray binaries with use of evolutionary calculations.
The calculated luminosities are compared with observed ones in terms of time-averaged mass accretion rate. 
It is shown that neutrino emission by strong pion condensation can explain quiescent X-ray luminosity of SAX J1808.4-3658 and
we do not need direct Urca processes concerning nucleons and/or hyperons.

\end{abstract}

\date{\today}

\maketitle

%----------------------------------------------------------------------------
\section{Introduction}
 
 There have been reported observational results~\cite{Paradijs1987,Verbunt1994,Zavlin1996,Wijnands2005} of soft X-ray transients during quiescence in low mass X-ray binaries.
The emergent radiation flux may depend on the neutron star structure,
  which opens an important possibility to explore the internal structure and the
equation of state (EoS) of dense matter by comparing numerical results with the observations~\cite{Lattimer2007,Beznogov2015}.
Moreover, the  observations of  accreting neutron stars give valuable informations
to constrain the structure of neutron
stars~\cite{Chabrier1997,Yakovlev2003}.  The sites include cooling phase of
a neutron star during  X-ray burst period~\cite{Bahramian2014} and quiescence~\cite{Heinke2007}.
In particular, average accretion rates over the quiescent periods have been derived from outburst luminosities
and some quiescent luminosities have  been observed by  direct measurements of photons 
from the neutron star surface~\cite{Heinke2007}.

Old transiently accreting neutron stars in low mass X-ray binaries have been successively investigated~\cite{Yakovlev2003}.
Matter escapes from their low mass companions and accretes onto neutron stars.
The accreted matter is compressed under the weight of freshly accreted material and 
  the compressional heating is furthermore accompanied by the way of heating from deep inner layers~\cite{BisnobatyiKogan1970,HZ90}. 
Characteristic energy release from  the deep crust is in the range of 1-2~MeV per accreted nucleon. 
The accretion phenomena are considered  to be neither too long (months--weeks) nor too strong to heat up  the crust; 
  the internal equilibrium between the crust and the core could be maintained. 
The energy release due to compression (compressional heating)
and/or crustal heating could be rather strong to keep the neutron stars warm and as a result lead to the emission of  observed thermal radiation for X-ray transients~\cite{HZ0308}. 
The mean heating rate could be determined from the time-averaged mass accretion rate, where the averaging has to be performed over characteristic relaxation time $t >10^3$~yr. 
On the other hand, luminosities of X-ray transients have been studied by calculating their theoretical curves of accreting neutron stars in quiescent
states~\cite{Heinke2007,Beznogov2015}. To derive theoretical luminosities, i.e., (effective temperatures) an analytical relation  between
the effective and core temperatures has been adopted~\cite{Gudmundsson1983}. As a consequence, 
neutrino losses due to kaon and pion condensations become insufficient to explain the observational 
luminosities~\cite{Heinke2007,Heinke2009}. Although the adopted relation between the two temperatures is convenient
to construct models of transient luminosities, the analytical relation should be examined
from the point of stellar evolution, where compressional heating is included. 
%Therefore, it is worthwhile to study the transient luminosities in detail with use of a stellar evolution code.

In the present study, our aim is to examine the relation between the luminosity of transiently accreting neutron stars and the time-averaged mass accretion rate using our spherically symmetric stellar evolutionary code. 
The time-averaged accretion rate is taken into account during the quiescent era, because
it becomes possible to compare the full calculations with simplified ones;
both the inner structure of the neutron star and the surface composition are combined to compare the luminosities observed from X-ray transients.
In \S~\ref{sec:method}, we present the basic equations and input physics.  We construct the quiescent neutron star models in \S~\ref{sec:luminosity}.
Our results of luminosities with use of  steady states under the constant mass accretion rates 
are presented in \S~\ref{sec:result}. 
Discussion is given in \S~\ref{sec:discussion}.

\section{Basic Equations and Physical Inputs}
\label{sec:method}

The general relativistic evolutionary equations of a spherical star in hydrostatic equilibrium are  formulated as follows~\cite{Thorne1977},
%%%%%%%%%%%%%%%%%%%%%%%%%%%%%%%%%%%%%%%%%%%%%%%%%%%%%%%%%%%%%%%%%%%%%%%%%%%%%%%%%%%%%%%%%%%%%%%%%%%%%%%%%%%%%%%%
%%%% Basic Equations %%%%%%%%%%%%%%%%%%%%%%%%%%%%%%%%%%%%%%%%%%%%%%%%%%%%%%%%%%%%%%%%%%%%%%%%%%%%%%%%%%%%%%%%%%%%%%
%%%%%%%%%%%%%%%%%%%%%%%%%%%%%%%%%%%%%%%%%%%%%%%%%%%%%%%%%%%%%%%%%%%%%%%%%%%%%%%%%%%%%%%%%%%%%%%%%%%%%%%%%%%%%%%%
\begin{eqnarray}
   \frac{\partial M_{tr}}{\partial r} & = & 4\pi r^{2} \rho~, \label{eq:1} \\
   \frac{\partial P}{\partial r} & = & -\frac{GM_{tr}\rho}{r^{2}}
     \left(1+\frac{P}{\rho c^{2}}\right)
     \left(1+\frac{4\pi r^{3}P}{M_{tr}c^{2}}\right)
     {V}^{2}~, \label{eq:2} \\
   %  
   %\frac{\partial (L_{r}e^{2\phi/c^{2}})}{\partial M_{r}} & = & 
   %  e^{2\phi/c^{2}}\left(\varepsilon_{\rm n}-\varepsilon_{\nu}-e^{-\phi/c^{2}}
   %  T\frac{\partial s}{\partial t_{\infty}}\right)~, \label{eq:3} \\
   \frac{\partial (L_{r}e^{2\phi/c^{2}})}{\partial M_{r}} & = & 
     e^{2\phi/c^{2}}\left(\varepsilon_{\rm n}-\varepsilon_{\nu}+\varepsilon_{\rm g}\right)~, \label{eq:3} \\
   %  
   %\frac{\partial \ln T}{\partial \ln P} & = & {\rm min}(\nabla_{\rm rad}, \nabla_{\rm ad})~, \label{eq:4} \\
   \frac{\partial \ln T}{\partial \ln P} & = & \nabla_{\rm rad}~, \label{eq:4} \\
   %
   %%%%e^{-\phi/c^{2}}\frac{\partial Y_{i}}{\partial t_{\infty}} & = & \alpha_{i} \label{eq:5} \\
   %e^{-\phi/c^{2}}\frac{\partial Y_{i}}{\partial t_{\infty}} & = & {\rm [Right~hand~side~of~the~rate~euation~(\ref{eq:rateeq})]} \label{eq:5} \\
   %
   \frac{\partial \phi}{\partial M_{tr}} & = & \frac{G(M_{tr}+4\pi r^{3}P/c^{2})}
     {4\pi r^{4}\rho}{V}^{2}~, \label{eq:5} %
\end{eqnarray} 
where 
\begin{eqnarray}
   \frac{\partial M_{tr}}{\partial M_{r}} & = & \frac{\rho}{\rho_0}
     {V}^{-1} ,\, \, \,  {V}\equiv \left(1-\frac{2GM_{tr}}{c^{2}r}\right)^{-1/2}. \label{eq:v}
\end{eqnarray}
%%%%%%%%%%%%%%%%%%%%%%%%%%%%%%%%%%%%%%%%%%%%%%%%%%%%%%%%%%%%%%%%%%%%%%%%%%%%%%%%%%%%%%%%%%%%%%%%%%%%%
We define the quantities used above set of equations in the followings; 
$r$: circumferential radius,
$\rho_0$: rest mass density, 
$\rho$: total mass-energy density, 
$T$: local (non red-shifted) temperature, 
$P$: pressure,
$M_{r}$: baryonic  mass inside the radius $r$, 
$M_{tr}$: gravitational mass inside the radius $r$, 
$\phi$: gravitational potential, 
%$s$: specific entropy, 
%$t_{\infty}$: Schwarzschild time coordinate (proper time at a distant observer), 
$L_r$: local luminosity, 
$\varepsilon_{\rm n}$: heating rate by nuclear burning, 
$\varepsilon_{\nu}$: cooling rate by escaping neutrinos,
$\varepsilon_{\rm g}$: gravitational energy release,
%, $\alpha_{i}$: nuclear reaction rate for the $i$-th particle. 
$\nabla_{\rm rad}$: so called {\it radiative temperature gradient} in which both the radiative opacity and electron conduction 
can be included together.
%have been included since they have the same functional form.

In the accretion layer, 
   the mass fraction coordinate with changing mass $q$ [=$M_r/M(t)$] is utilized, 
   which is the most suitable method for computations of stellar structure when the total stellar mass $M$ varies \cite{Sugimoto1981}.
As a consequence, the gravitational energy release $\varepsilon_{\rm g}$ is divided into two parts~\cite{Fujimoto1984}:
\begin{eqnarray}
  \varepsilon_{\rm g}^{\rm (nh)} &=& 
      -\exp\left( -\frac{\phi}{c^2}\right) 
        \left( T \left. \frac{\partial s}{\partial t} \right|_q + \mu_i \left. \frac{\partial N_i}{\partial t} \right|_q \right),  \label{eq:eg_nh}\\
  \varepsilon_{\rm g}^{\rm (h)}  &=& 
        \exp\left( -\frac{\phi}{c^2}\right) \dot{M} 
        \left( T \left. \frac{\partial s}{\partial \ln q} \right|_t + \mu_i \left. \frac{\partial N_i}{\partial \ln q} \right|_t \right), \label{eq:eg_h}
\end{eqnarray}
where
  $s$ is specific entropy, 
  $t$ is Schwarzschild time coordinate,
  and $\dot{M}$ is mass accretion rate.
  $\mu_i$ and $N_i$ are, respectively, chemical potential and number per unit mass of the $i$-th elements.
Equation (\ref{eq:eg_nh})  is so-called {\it nonhomologous} term and 
    equation (\ref{eq:eg_h}) is {\it homologous} term of compressional heating.

The radiative zero boundary condition is imposed at the outer boundary.  An outermost mesh-point, which is close enough to the photosphere
for our investigation, is given at $q=1-9.9\times10^{-20}$. %\cite{Fujimoto1984}. 

The above set of  equations (\ref{eq:1}) -- (\ref{eq:5}) can be solved numerically with use of the Henyey-type numerical scheme of implicit method.
We adopt the evolution code of a spherically symmetric neutron star \cite{Hanawa1984,Fujimoto1984}.

%%%%%%%%%%%%%%%%%%%%%%%%%%%%%%%%%%%%%%
For EoS concerning outer layers of the neutron star ($\rm \rho < 5\times 10^7~g~cm^{-3}$), an ideal gas plus radiation is
adopted with the electron degeneracy and the Coulomb liquid correction included \cite{Slattery1980}. 
For the inner layers of $\rm \rho < 6\times 10^{12}~g~cm^{-3}$, 
EoS has been adopted from Ref.~\cite{Richardson1982}.
For the further inner layers,
EoS constructed by Lattimer and Swesty~\cite{Lattimer1991} (hereafter referred to LS) is used with the incompressibility of 220 MeV
under the constraint of $\beta$-equilibrium. Furthermore, we include the effects of pion condensations~\cite{Umeda1994}. 
These effects result in softening of the EoS  and 
the maximum mass of the neutron star is reduced from $2.53$ to $2.21~M_{\odot}$.  
Nevertheless, it is safe to
exceed  the recent observations of both
$1.97\pm0.04~M_{\odot}$~\cite{Demorest2010} and $2.0\pm0.04~M_{\odot}$~\cite{Antoniadis2013} for a 1$\sigma$ level.
It should be noted that  EoS constructed by LS has been still used  for the simulations of
supernova explosions~\cite{Couch2013,Couch&O'Connor2014,Couch&Ott2015}. 
While EoS above the nuclear density ($\rho_0$) is very uncertain,  LS has been constructed on the basis of 
detailed micro-physics below the nuclear density. On the other hand, EoS  by Shen et al.~\cite{Shen1998} seems to be too stiff to
induce the supernova explosions~\cite{Suwa+2013} and there are some challenges to soften this EoS by including hyperons~\cite{Ishizuka2008}.
In any case, EoS of supernova matter should be  also applied to the neutron star matter~\cite{Couch2013c}.

Neutrino emissivities include bremsstrahlung of nucleon-nucleon, modified Urca \cite{Friman1979},
 and electron-ion bremsstrahlung~\cite{Festa1969}, and electron-positron pair, photo, and plasmon processes \cite{Beaudet1967}. 
 We also include the strong neutrino losses due to pion condensations by  Muto et al.~\cite{Muto1993}, where pion condensation
 begins at about $4\times10^{14}~\rm g~cm^{-3}$ (see \S~\ref{sec:luminosity}). Although the short range correlation in nuclei is
very uncertain, recent experiments have indicated that the pion condensation
 begins  at $(1.9\pm 0.3)\rho_0$~\cite{Yako+2005} and furthermore $(1.8 - 2.4)\rho_0$~\cite{Ichimura+2006}, which support the prediction of 
  Muto et al.~\cite{Muto1993}.

%%%%%%%%% Figure 1 and 2 %%%%%%%%%%%%%%%%%%%%
\begin{figure}[tp]
\begin{center}
    \begin{tabular}{lcr}
    \begin{minipage}[t]{0.5\hsize} 
       \mbox{\raisebox{-1.7mm}{\includegraphics[scale=1.20]{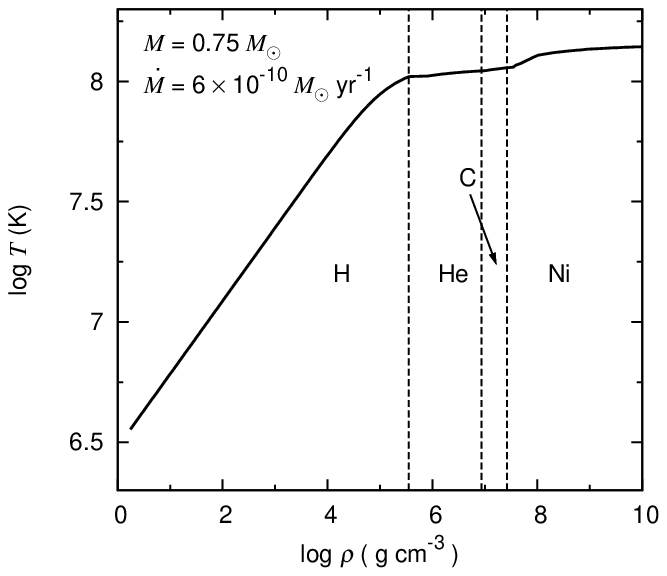}}}
       \caption{
           Model  for $M = 0.75~M_{\odot}$ and $\dot{M} = 6\times10^{-10}~M_{\odot}~\rm yr^{-1}$ .
           Light elements are included in the accretion layers. The dashed lines indicate the boundaries of  these elemental species.
     }
     \label{fig:init}
    \end{minipage}
    &\,&
    \begin{minipage}[t]{0.5\hsize}
        \mbox{\includegraphics[scale=1.20]{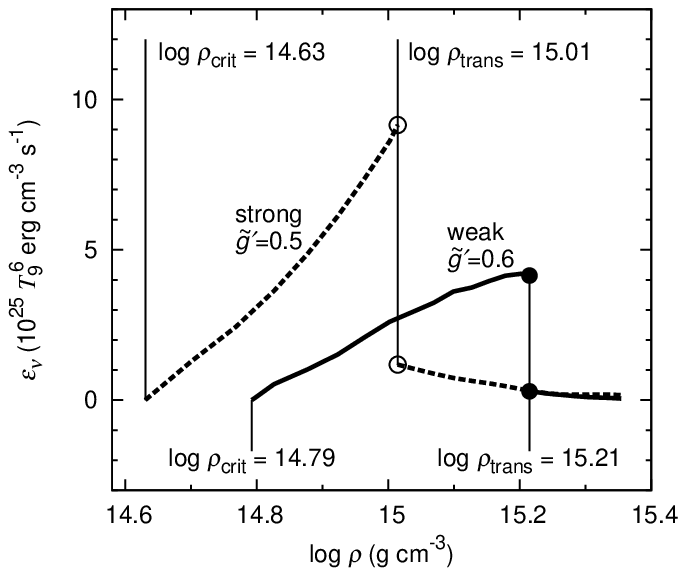}}
        \caption{
            Neutrino emission rates for $\Tilde{g}' = 0.5$ and $\Tilde{g}' = 0.6$~\cite{Muto1993}.
            The vertical lines indicate the density at the appearance of the condensed states for either $\pi^{c}$ ($\rho_{\rm crit}$) or $\pi^{0} \pi^{c}$ ($\rho_{\rm trans}$). 
            The dashed and solid curves correspond to $\Tilde{g}' = 0.5$ which causes strong cooling and $\Tilde{g}' = 0.6$ weak cooling, respectively.
            The transition points ($\rho_{\rm trans}$) are shown by the filled or open circles.
        }
        \label{fig:neutrino}
    \end{minipage}
    \end{tabular}
\end{center}
\end{figure}

%%% updated opacity
Considering the significant progress in numerical calculations of opacities, we have updated the opacities in Ref.~\cite{Fujimoto1984} as follows.
We adopt  the electron conductive opacity~\cite{Potekhin2015} without magnetic field in both the ocean and the crust.
We use the proton charge profile inside a nucleus~\cite{Oyamatsu1993}. 
We also adopt the electron conductive opacity~\cite{Baiko2001} in the core
where we include only electron-proton scattering opacity because this process is dominant for the realistic dense matter
    if we neglect the effect of superfluidity.
We take into account radiative opacities contributed from free-free \cite{Schatz1999} and relativistic electron scatterings \cite{Paczynsk1983}.
However, these updates do not significantly change our results.

In the present paper, we include the crustal heating \cite{HZ90}, 
\begin{equation}
    Q_i=6.03 \times \dot{M}~q_i~ 10^{43}~ \rm erg~s^{-1}~, \label{eq:crustheat}
\end{equation}
where $q_i$ represents the effective heat release per unit nucleon on $i$-th reaction surface (see Tables 1 and 2 in Ref.~\cite{HZ90})
and $\dot{M}$  represents the mass accretion rate in units of $M_{\odot}$ yr$^{-1}$.
The energy generation rate $\varepsilon_{\rm n}$ can be evaluated from $Q_i / \Delta M$.  
The mass $\Delta M$ corresponds to the range of the  Lagrange mass coordinate in which the $i$-th reaction surface locates.

As for the surface composition in the accretion layer, we examine two cases: Ni only and Ni, C, He and H (hereafter designated
as light elements).
For example, Fig.~\ref{fig:init} shows a model  for the neutron star mass $0.75~M_{\odot}$ and 
$\dot{M}= 6\times10^{-10}~M_{\odot}~\rm yr~^{-1}$  with the surface composition of light elements.
%Light elements in addition to $^{56}$Ni,  $^1$H and $^4$He and $^{12}$C are included in the accretion layers (hereafter designated
%'light elements' or 'H'). 
The dashed lines indicate the boundaries of individual  elements. 
Note that we do not include thermonuclear burning of light elements even if we use the models with the surface composition of light elements
because we consider only quiescent neutron stars. Thus the energy generation rate $\varepsilon_{\rm n}$ is only given by the crustal heating.

To illustrate the effects of pion condensations, we show the neutrino emission rates $\epsilon_{\nu}$ due to the pion condensations as a function of density in Fig.~\ref{fig:neutrino},
    where the emission rates are adopted from Muto et al.~\cite{Muto1993};
They found that the appearance of $\pi^c$-condensed state  begins at $\rho_{\rm crit}$ and the combined $\pi^0 \pi^c$-condensed state at $\rho_{\rm trans}$. 
The transition points ($\rho_{\rm trans}$) are indicated by the filled or open circles in Fig.~\ref{fig:neutrino}.
The neutrino emission rates and the density of the phase transition depend on the dimensionless parameter $\Tilde{g}'$.
We adopt the case  $\Tilde{g}' = 0.5$ in the evolutionary calculations which causes larger neutrino emission as seen from Fig.~\ref{fig:neutrino}
and in the present study we
regard the cooling as exotic.  We note that  the cooling is much more efficient to determine the structure
of the neutron star with enough mass for  $\pi^c$-condensed state compared to the $\pi^0 \pi^c$-condensed state. This is because
the neutron star contains the density region where the exotic cooling due to the  $\pi^c$-condensed state is  dominant
as seen in Fig.~\ref{fig:neutrino}. 

%%%%%%%%% Figure 3 %%%%%%%%%%%%%%%%%%%%%%%%%%%%%%%
\begin{figure}[tp]
    \begin{center} 
    \begin{tabular}{ccc} \hspace{-2ex}
    \begin{minipage}[c]{0.5\hsize}
        \includegraphics[scale=1.20]{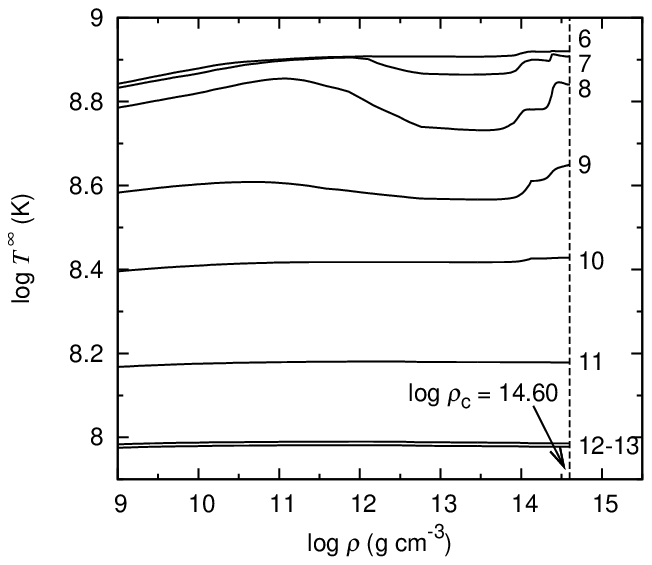}
    \end{minipage}
    &\,&
    \begin{minipage}[c]{0.5\hsize}
        \includegraphics[scale=1.20]{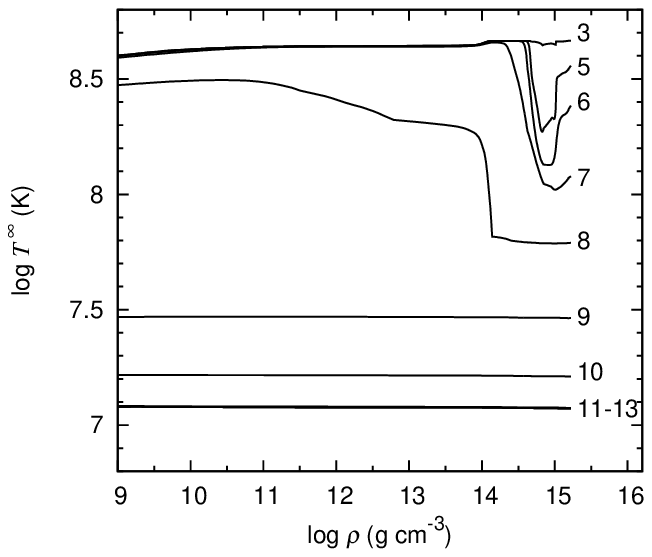}
    \end{minipage}
    \end{tabular}
    \caption{
         Time evolution of red-shifted temperatures towards the steady state of neutron stars.
          Note that the red-shifted temperature is defined by $T^{\infty} \equiv \exp(\phi/c^2) T$.
         The left and right panels show the results for $M=0.75~M_{\odot}$ and $M = 2.01~M_{\odot}$, respectively.
         The accretion rates are $1\times10^{-10}~M_{\odot}~\rm yr~^{-1}$, and the surface compositions are Ni.
         The numerals attached in each curve from 3 to 13 indicate a series of  $\log~t$.
        The time $t$ in units of second  is measured from the beginning of the computation.
    }
    \label{fig:evol_temp}
\end{center}
\end{figure}

\section{Thermal Structures in the Quiescent States}
\label{sec:luminosity}

%%%%%%%%% Figure 4, 5 %%%%%%%%%%%%%%%%%%%%%%%%%%%%%%%
\begin{figure}[tp]
\begin{center}
    \begin{tabular}{ccc} %\hspace{-2ex}
    \begin{minipage}[t]{0.5\hsize}
        \includegraphics[scale=1.2]{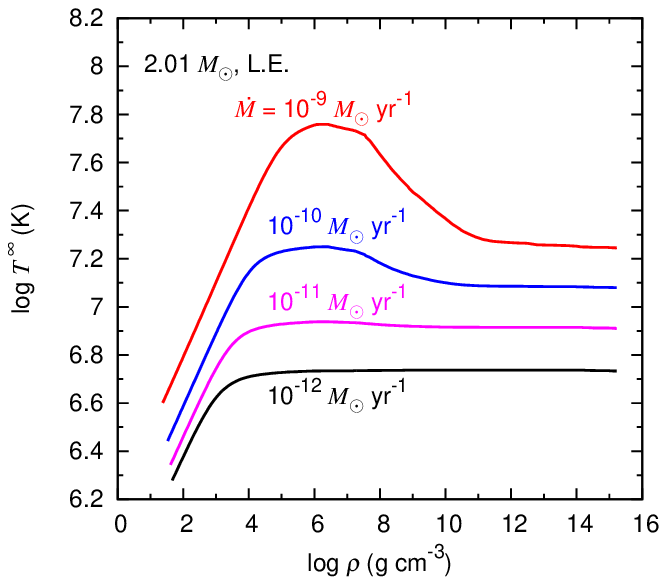}
        \caption{ 
            Temperature vs. density in the steady states with the compressional heating.
            Shown are the results of models for $M = 2.01 M_{\odot}$ with use of the surface composition of light elements (designated by L.E.).
            The mass accretion rates are attached in each curve.
        }
        \label{fig:rho-T_LE}
    \end{minipage}
    &\,&
    \begin{minipage}[t]{0.5\hsize}
         \includegraphics[scale=1.2]{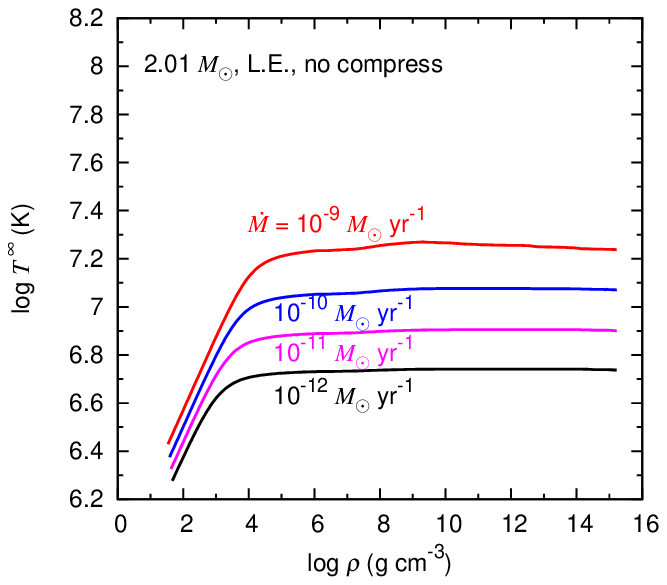}
         \caption{
            Same as Fig.~\ref{fig:rho-T_LE} but for the compressional heating neglected.
        }
        \label{fig:rho-T_LE_nocmp}
    \end{minipage}
    \end{tabular}
\end{center}
\end{figure}

We perform the evolutionary calculations of accreting neutron stars with the continuous accretion ($\dot{M}=$ constant)
    until the the nonhomologous part of the gravitational energy release vanishes \cite{Fujimoto1984}.
The left panel in Fig.~\ref{fig:evol_temp} shows the variation of the thermal structure during $t=10^6 - 10^{13}$~s for $M=0.75~M_{\odot}$
    with $\dot{M}= 1\times10^{-10}~M_{\odot}~\rm yr~^{-1}$.  
We can recognize the temperature distribution attains the steady state for $t\geq 10^{12}$~s. 
Since the central density ($\rho_{\rm c} = 10^{14.60}~{\rm g~cm^{-3}}$) does not reach the critical density ($\rho_{\rm crit}=10^{14.63}~{\rm g~cm^{-3}}$),
    cooling due to the pion condensations does not appear.
The right panel in Fig.~\ref{fig:evol_temp} shows the variation of the thermal structure during $t=10^3 - 10^{13}~{\rm s}$ for $M=2.01~M_{\odot}$
    with $\dot{M}= 1\times10^{-10}~M_{\odot}~\rm yr~^{-1}$. 
One sees the significant effects of the cooling around $\rho > 10^{14} ~{\rm g~cm^{-3}}$ 
    during $t=10^3 - 10^{8}$~s. We recognize that the temperature distribution attains the steady state for $t\geq 10^{11}$~s.  
We find that the exotic cooling becomes appreciable in the neutron star of $M \geq 0.92~M_{\odot}$, because the central density exceeds $\rho_{\rm crit}$.

Figure~\ref{fig:rho-T_LE} shows the thermal structures in the steady states for models of $2.01~M_{\odot}$ with the surface composition of light elements.
These models have the peak temperatures at the density $\rho \sim 10^{6.4}~{\rm g~cm^{-3}}$
  which indicate the boundaries concerning the thermal flow towards either an inner or an outer region.
These peaks are due to the compressional heating $\varepsilon_{\rm g}^{\rm (h)}$ of Eq.~(\ref{eq:eg_h}).
They increase the effective temperatures significantly.
However, these steady states with the constant accretion do not correspond to the observed quiescent states of neutron stars
   because the accretion rates in the observed quiescent phase are much smaller than those of our calculations.

To compare the obtained luminosities with the those of previous studies~\cite{potekin1997,Yakovlev2003,Yakovlev2004,Beznogov2015} which do not include the compressional heating,
    we perform the calculations without the compressional heating until the steady states are archived.
Figure~\ref{fig:rho-T_LE_nocmp} shows the thermal structures in the steady states without the compressional heating.
The peaks of the temperature disappear due to the lack of the compressional heating.

%%%%%%%%% Figure 6, 7 %%%%%%%%%%%%%%%%%%%%%%%%%%%%%%%
\begin{figure}[tp]
\begin{center}
   \begin{tabular}{ccc} %\hspace{-2ex}
   \begin{minipage}[t]{0.5\hsize}
        \includegraphics[scale=1.2]{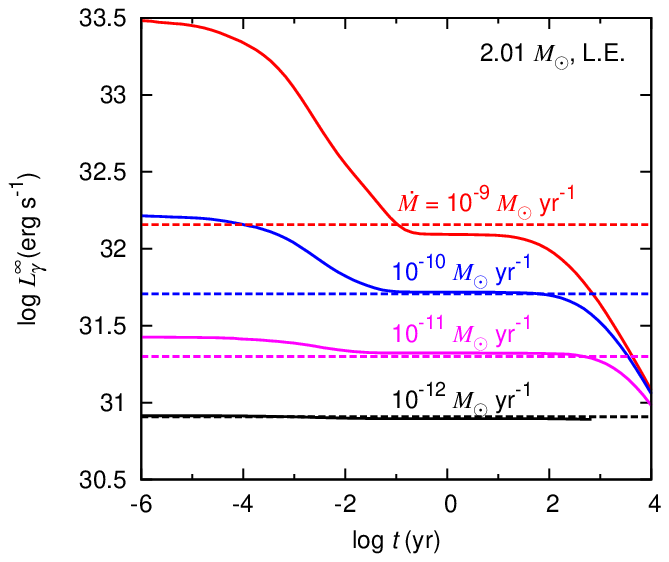}
        \caption{
            Solid curves represent the time evolutions of luminosities after the mass accretion rates are set to be zero.
            The mass accretion rates just before the cooling begins are indicated for each curve.
            Dotted lines represent the luminosities in the steady states constructed without the compressional heating.
        }
        \label{fig:Levol}
    \end{minipage}
    &\,&
    \begin{minipage}[t]{0.5\hsize} 
        \includegraphics[scale=1.2]{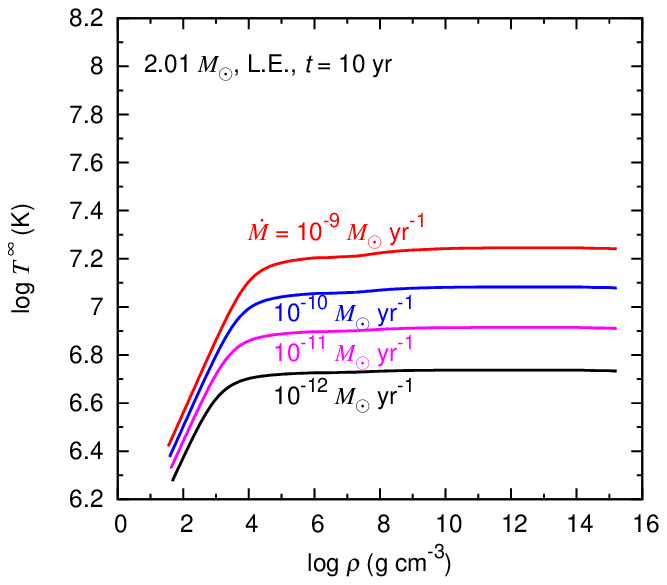}
        \caption{
            Temperature vs. density at $10~{\rm yr}$ from the end of the accretion,
            where the mass accretion rates are indicated in each curve.
        }
        \label{fig:rho-T_LE_10yr}

    \end{minipage}
    \end{tabular}
\end{center}
\end{figure}

Now, we need to examine whether the steady states without the compressional heating correspond to the observed quiescent states.
We calculate the evolutions without the accretion ($\dot{M} = 0$), where the initial models are the steady state models constructed under the constant accretion.
The results are shown in Figs.~\ref{fig:Levol} and \ref{fig:rho-T_LE_10yr}.
The solid curves in Fig.~\ref{fig:Levol} show the time evolutions of the luminosities, where the horizontal axis is the elapsed time measured from the end of accretion.
We find that the luminosities decrease significantly during $0.1~{\rm yr}$ and they become constant in around $100~{\rm yr}$.
As a result, we can regard the states with constant luminosities as the quiescent states,
    because the periods of the constant luminosities are longer than the typical periods of the observed quiescent states.
Figure \ref{fig:Levol}  shows also the luminosities (dotted lines) in the steady states without the compressional heating.
The luminosities without the compressional heating are exactly similar to the luminosities in the evolutionary calculations during $t = 0.1 - 100~{\rm yr}$.
Moreover, these two models have almost the same thermal structures (see Figs.~\ref{fig:rho-T_LE_nocmp} and \ref{fig:rho-T_LE_10yr}),
    which indicates that the heat due to the compressional heating is radiated away from the neutron star in less than $0.1~{\rm yr}$.
As a consequence, the steady state models without the compressional heating correspond to the observed quiescent neutron star.

%%%%%%%%% Figure 8 %%%%%%%%%%%%%%%%%%%%%%%%%%%%%%%
\begin{figure}[tp]
\begin{center}
    \begin{tabular}{ccc} \hspace{-2ex}
        \begin{minipage}[c]{0.50\hsize}
            \includegraphics[scale=1.2]{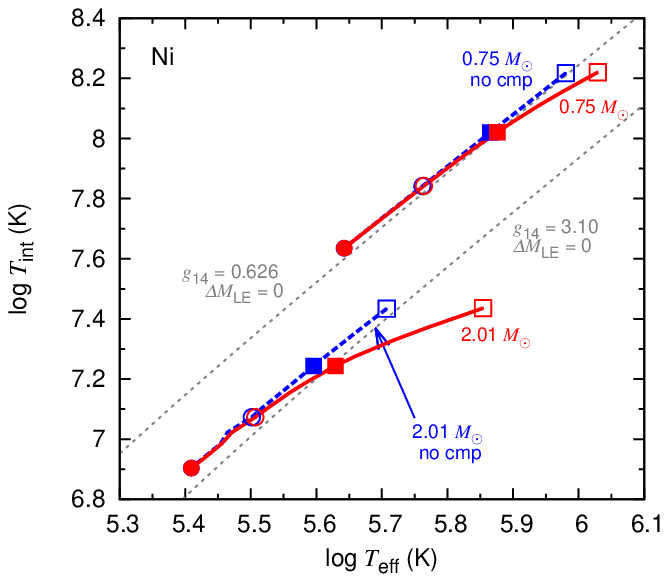}
        \end{minipage}
        &\,&
        \begin{minipage}[c]{0.50\hsize}
            \begin{center}
                \includegraphics[scale=1.2]{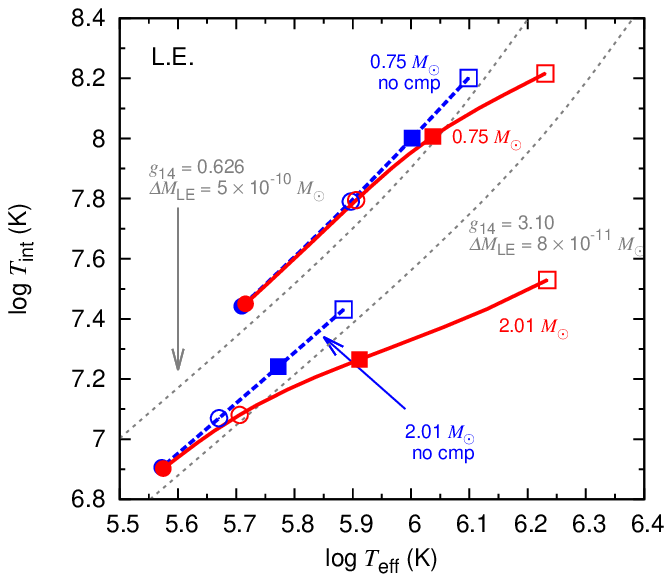}
            \end{center}
        \end{minipage}
    \end{tabular} 
    \caption{
         Relations between the effective and the inner temperatures for $M = 0.75$ and $2.01~M_{\odot}$,  
             where the inner temperature is evaluated at the position of $\rho = 10^{10}~{\rm g~cm^{-3}}$.
         The left and right panels show the results for the surface composition of Ni and light elements (L.E.), respectively.
         The solid curves are the results with compressional heating and
             the dashed lines (`no cmp') represent the models without the compressional heating.
         The filled circles, open circles, filled squares, open squares, and filled triangles are, respectively, 
         the results for $\dot{M} = 10^{-12}, 10^{-11}, 10^{-10},10^{-9} ~M_{\odot}~{\rm yr^{-1}}$.
         The thin dotted lines are obtained from an analytical formula in Ref.~\cite{potekin1997},
             where $g_{14}$ is the surface gravity in units of $10^{14}~{\rm cm~s^{-2}}$ and $\Delta M_{\rm LE}$ indicates the mass of accreted light elements.
    }
    \label{fig:tstb}
\end{center}
\end{figure}

Figure~\ref{fig:tstb} shows relations between the effective and the inner temperatures for $M = 0.75$ and $2.01~M_{\odot}$ without the surface composition of Ni and light elements (L.E.).  
The inner temperature is evaluated at the point of $\rho = 10^{10}~{\rm g~cm^{-3}}$ because
the temperature does not change appreciably above this density until the central region,
where crustal heating is deposited gradually toward the center~ \cite{Fujimoto1984}. 
The surface gravity $g_{14}$ is expressed in units of $10^{14}~{\rm cm~s^{-2}}$, where $g = GM / (R^2 \sqrt{1-r_g/R}$) and $r_g$ is the Schwarzschild radius.
The values are $g_{14} = 0.626$ and $3.10$ for the models with $M = 0.75~M_{\odot} ~(R = 13.2~{\rm km})$ and $M = 2.01~M_{\odot} ~(R = 11.2~{\rm km})$, respectively.
We note that both temperatures are non red-shifted temperatures~\cite{potekin1997}.
Our results without the compressional heating are qualitatively consistent with those in  Ref.~\cite{potekin1997}. 
However, if we include the compressional heating, the results deviate as seen in Fig.~\ref{fig:tstb}.
The compressional heating causes the higher effective temperatures compared with those of the previous study.
We conclude that the models without the compressional heating correspond to the previous models~\cite{potekin1997,Yakovlev2003,Yakovlev2004,Beznogov2015}.

\renewcommand{\arraystretch}{1.25}  
\begin{table}[tn]
   \centering
   \caption{
        Logarithm of red-shifted luminosities in units of ${\rm erg \ s^{-1}}$ for calculated models without the compressional heating. 
        The gravitational masses and circumferential radii are given for reference. 
   }
   \scalebox{1.00}[1.00]{
   {\tabcolsep = 2.7mm
   \begin{tabular}{c ccc c ccc} \hline \hline
      \multirow{3}{*}{$\dot{M}~(M_{\odot}~{\rm yr^{-1}})$} & \multicolumn{3}{c}{Ni}  &
                                                                             &  \multicolumn{3}{c}{Light elements}  \\ \cline{2-4} \cline{6-8}
         &0.75~$M_{\odot}$  &1.40~$M_{\odot}$  &2.01~$M_{\odot}$  &     
            &0.75~$M_{\odot}$  &1.40~$M_{\odot}$  &2.01~$M_{\odot}$  \\ 
         &13.2 km&12.9 km&11.2 km& & 13.2 km&12.9 km& 11.2 km \\ \hline
      $1.0 \times 10^{-12}$   &31.59  &30.33  &30.26   &     &31.86   &30.96   &30.91    \\
      $1.5 \times 10^{-12}$   &31.69  &30.40  &30.33   &     &32.03   &31.04   &30.96    \\
      $3.0 \times 10^{-12}$   &31.84  &30.50  &30.44   &     &32.28   &31.17   &31.09    \\
      $5.0 \times 10^{-12}$   &31.94  &30.59  &30.48   &     &32.44   &31.26   &31.18    \\
      $1.0 \times 10^{-11}$   &32.07  &30.70  &30.63   &     &32.61   &31.38   &31.30    \\
      $1.5 \times 10^{-11}$   &32.15  &30.76  &30.69   &     &32.70   &31.46   &31.38    \\
      $3.0 \times 10^{-11}$   &32.27  &30.88  &30.81   &     &32.83   &31.58   &31.50    \\
      $5.0 \times 10^{-11}$   &32.36  &30.96  &30.89   &     &32.92   &31.67   &31.59    \\
      $1.0 \times 10^{-10}$   &32.49  &31.08  &31.01   &     &33.03   &31.79   &31.71    \\
      $1.5 \times 10^{-10}$   &32.56  &31.16  &31.08   &     &33.10   &31.87   &31.78    \\
      $3.0 \times 10^{-10}$   &32.69  &31.29  &31.20   &     &33.21   &32.00   &31.91    \\
      $5.0 \times 10^{-10}$   &32.79  &31.40  &31.30   &     &33.29   &32.11   &32.01    \\
      $1.0 \times 10^{-9} $    &32.95  &31.58  &31.45   &     &33.42   &32.28   &32.16    \\ \hline \hline
   \end{tabular}
   }
   }
   \label{tab:numedata}
\end{table}

\renewcommand{\arraystretch}{1}
\section{Results}
\label{sec:result}

%%%%%%%%% Figure 7 %%%%%%%%%%%%%%%%%%%%%%%%%%%%%%%
\begin{figure}[tp]
    \vspace{-10eM}
    \begin{center}
        ~\hspace{3eM} \includegraphics[scale=1.5,clip]{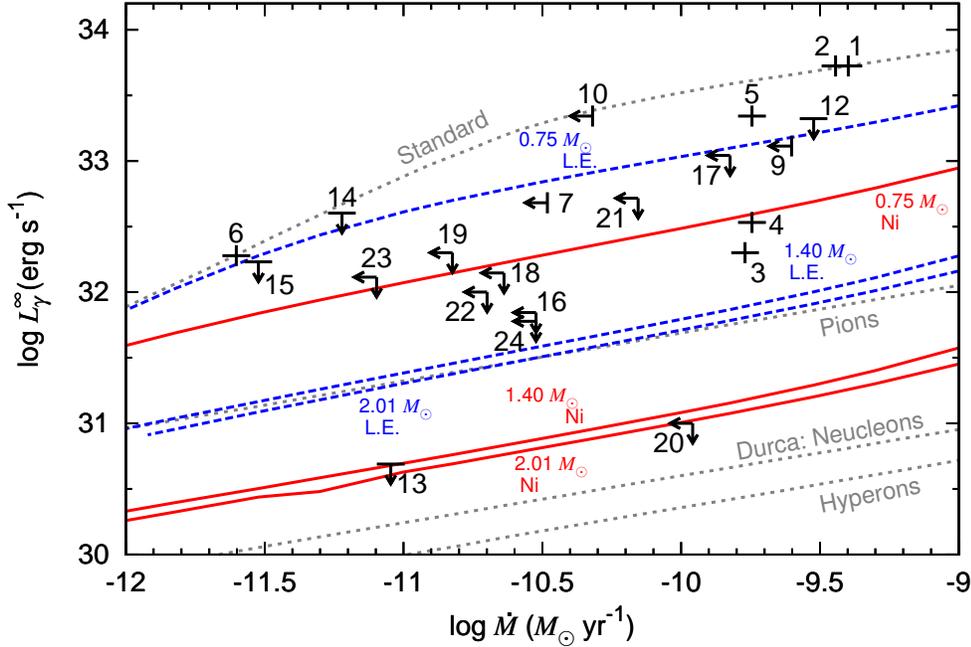}
        \caption{
            Observed (red-shifted) luminosity vs. time-averaged mass accretion rate for $M = 0.75$, $1.40$, and $2.01~M_{\odot}$. 
            Solid and dashed curves correspond to the models with the surface composition of  Ni and light elements (L.E.), respectively.
           We show the results of the steady states without the compressional heating.
           The thin dotted curves are taken from Yakovlev et al.~\cite{Yakovlev2004}. 
           The crosses and arrows indicate the observational values and their upper limits of the observations, respectively~\cite{Heinke2007,Heinke2009}.
           The numerals from `1' to `24' correspond to the following objects; 
               1: Aql X-1,
               2: 4U 1608-522,
               3: MXB 1659-29,
               4: NGC 6440 X-1,
               5: RX J1709-2639,
               6: IGR 00291+5934,
               7: Cen X-4,
               8: KS 1731-260,
               9: 1M 1716-315,
              10: 4U 1730-22,
              11: 4U 2129+47,
              12: Terzan 5,
              13: SAX J1808.4-3658,
              14: XTE J1751-305,
              15: XTE J1814-338,
              16: EXO 1747-214,
              17: Terzan 1,
              18: XTE J2123-058,
              19: SAX J1810.8-2609,
              20: 1H 19605+000,
              21: 2S 1803-45,
              22: XTE J0929-314,
              23: XTE J1807-294,
              24: NGC 6440 X-2.
            The data for each object are taken from Ref.~\cite{Beznogov2015}.
            We do not insert the data of '8' and '11' because these objects cannot constrain the neutrino process
                due to the lack of the information about the mass accretion rates.
        }
        \label{fig:luminosity}
    \end{center}
\end{figure}

The red-shifted luminosities ($L^{\infty}_{\gamma}$) of our neutron star models for $M = 0.75 - 2.01~M_{\odot}$ without the compressional heating are given in 
Table~\ref{tab:numedata} for the surface composition of Ni and light elements.
In the case of light elements, we have lower opacity and consequently higher luminosity than in the case of Ni. 
Figure~\ref{fig:luminosity} shows red-shifted luminosity against the accretion rate.
The numbers `1-24' designate the observational data and the associated
arrows indicate the upper limits of individual  data for X-ray transients~\cite{Beznogov2015}. 
The thin dotted curves are taken from Yakovlev et al.~\cite{Yakovlev2003}, where the data of `13' cannot be
explained without inclusion of additional strong cooling processes due to direct Urca processes involving nucleons and/or 
hyperons~\cite{Heinke2009,Beznogov2015}. 

On the other hand, our computational results are shown in Fig.~\ref{fig:luminosity} for $M = 0.75$, $1.40$ and  $2.01~M_{\odot}$.
The solid curves correspond to the results with the surface composition of  Ni and the dashed curves stand for light elements (L.E.). 
As a whole, we can obtain reasonable agreement by comparing our results with the observed data. However,
we must choose the mass less than $0.75~M_{\odot}$ even if we take the surface composition of light elements to explain the data of `1', `2', and `10'. 
In particular, it should be noted that the observational data of `13' could be explained
  by using the cooling rate by Muto et al.~\cite{Muto1993}  without introducing additional/exotic cooling mechanism. 
In other words, neutron stars whose masses are larger than $1.40~M_{\odot}$  with the surface composition of Ni could explain the data of `13'. 

\section{Discussion}
\label{sec:discussion}

As seen in Fig.~\ref{fig:luminosity}, luminous observational data of `1', `2', and `10' cannot
be fitted by adopting $M = 0.75~M_{\odot}$ with the surface composition of light elements. 
Compared to the conventional value of the neutron star mass $1.40~M_{\odot}$, it may be too light  to be observed. 
Moreover, the neutron star mass is estimated to be $0.75-1.28~M_{\odot}$ for the observational data `13' 
which is identified to be the X-ray transient SAX J1808.4-3658~\cite{Elebert2009}. 
Since only the upper limit of the luminosity is observed, neutrino emission rate by Muto et al.~\cite{Muto1993} is barely consistent with 
the observation  `13'.
We must note that results by Muto et al.~\cite{Muto1993} has large uncertainty concerning the nuclear interactions.
Moreover, if we include the effects of superfluidity which tend to cancel the effects of neutrino emissions, consistency between theory and observation may become difficult. 
Although critical temperature which induces superfluidity is very uncertain~\cite{Kaminker2006,Page2009,Shternin2011,Noda2013}, 
it should be investigated whether the consistency between computational results and observations could be maintained. 
Furthermore, we note that our models are inconsistent with the observations of isolated neutron star cooling.
We cannot account for the observations for our adopted EoS and neutrino emissions due to the pion condensation,
because the pion condensation occurs in all cases of $M > 0.75~M_{\odot}$
Inclusion of the effects of superfluidity may solve the problem and we will investigate them in the near future.

\begin{acknowledgements}
We thank Dr. Kenzo Arai for helpful discussion.
This work has been supported in part by a Grant-in-Aid for Scientific Research (24540278, 15K05083) of the Ministry of Education, Culture, Sports, Science and Technology of Japan.
\end{acknowledgements}

%---- BibTex ---   
%\bibliographystyle{h-physrev}
%\bibliography{database_kh01.20}
%-----

\end{document}